# The Degree Distribution of Random Birth-and-Death Network with Network Size Decline

Xiaojun Zhang*, Huilan Yang

School of Mathematical Sciences, University of Electronic Science and Technology of China,

Chengdu 611731, P.R. China

## Abstract

In this paper, we provide a general method to obtain the exact solutions of the degree distributions for RBDN with network size decline. First by stochastic process rules, the steady state transformation equations and steady state degree distribution equations are given in the case of $m \geqslant 3$, $0<p<1/2$ , then the average degree of network with $n$ nodes is introduced to calculate the degree distribution. Especially, taking $m=3$ as an example, we explain the detailed solving process, in which computer simulation is used to verify our degree distribution solutions. In addition, the tail characteristics of the degree distribution are discussed. Our findings suggest that the degree distributions will exhibit Poisson tail property for the declining RBDN.

* Corresponding author: Xiaojun Zhang Email address: sczhxj@uestc.edu.cn

# The Degree Distribution of Random Birth-and-Death Network with Network Size Decline

## 1. Introduction

In the real world, there exist many birth-and-death networks such as the World Wide Web [1-3], the communication networks [4-6], the friend relationship networks [7-10] and the food chain networks [11-13], in which nodes may enter or exit at any time. For this kind of evolving networks, the degree distribution is always one of the most important statistical properties. Several methods have been proposed to calculate their degree distributions like the first-order partial-differential equation method by Saldana [14], the mean-field approach [15] by Slater et al. [16], the rate equation approach by Sarshar and Roychowdhury [17] Moore et al. [18], Garcia-Domingo et al. [19] Ben-Naim and Krapivsky [20]. While these approaches aim at the networks whose size is growing or remain unchanged at each time step keep, Zhang et al. [21] put forward the stochastic process rules (SPR) based Markov chain method to solve evolving networks in which the network size may increase or decrease at each time step.

Furthermore, a random birth-and-death network model (RBDN) is considered [22], in which at each time step, a new node is added into the network with probability $p$ ($0< p <1$) with connection to $m$ old nodes uniformly, or an existing node is deleted from the network with the probability $q=1-p$. For network size decline ($0<p<1/2$), since $m$ is a critical parameter for the degree distributions, solution methods may vary for different $m$. While Zhang et al. [22] only consider a special case $m$=1, 2, in this paper, a general approach is proposed for solving the degree distributions of RBDN with network size decline in case of $m\geqslant 3$. Taking $m$=3 as an example, we provide the exact solutions of the degree distributions. Our findings also indicate that the tail of the degree distribution for the declining RBDN is subject to Poisson tail.

This paper is structured as follows: Section 2 gives the RBDN model with $0<p<1/2$ and its steady-state equations; Section 3 provides the method of solving degree distribution of RBDN with $0<p<1/2$; Section 4 further discusses the tail characteristics of the degree distribution and Section 5 concludes the paper.

## 2. Steady State Equations of RBDN with Network Size Decline

### 2.1 RBDN model

Consider the RBDN model [22]:

(i) The initial network is a complete graph with $m$+1 nodes, where $m$ is a positive integer;

(ii) At each unit of time, add a new node to the network with probability $p$ (0<$p$<1/2) and connect it with $m$ old nodes uniformly, or randomly delete a node from the network with probability $q$=1-$p$.

Note:

(a) Here we assume the low-bound of the network size $n_0 = 1$, that is, if the number of nodes in the network is $n_0$ at time $t$, then at time $t+1$, we only add a new node to the network with probability $p$ and connect it to the old node in the network.

(b) If at time $t$, a new node is added to the network and the network size is less than $m$, then the new node is connected to all old nodes.

### 2.2 Steady state transformation equations

Using SPR [21], we use $(n,k)$ to describe the state of node $v$, where $n$ is the number of nodes in the network that contains $v$, and $k$ is the degree of node $v$. Let $NK(t)$ denote the state of node $v$ at time $t$, the stochastic process $\{NK(t), t \ge 0\}$ is an ergodic aperiodic homogeneous Markov chain with the state space $E = \{(n,k), n \ge 1, 0 \le k \le n-1\}$.

Let $\tilde{\boldsymbol{P}}(t)$ be the probability distribution of $NK(t)$, i.e.

$$\tilde{P}_{(n,k)}(t) = P\{NK(t) = (n,k)\} \tag{1}$$

The state transformation equations are as follows (see the Appendix for the details):

$$\tilde{\boldsymbol{P}}(t+1) = \tilde{\boldsymbol{P}}(t) \cdot \boldsymbol{P} \tag{2}$$

where $\boldsymbol{P}$ is the one-step transition probability matrix. Let

$$\Pi_{(i,k)} = \lim_{t \to +\infty} P\{NK(t) = (i,k)\} \tag{3}$$

Taking the limit of Eq. (2) as $t \to +\infty$, the steady state transformation equations can be obtained

$$\begin{cases} \Pi_{(1,0)} = q\Pi_{(1,0)} + q\Pi_{(2,0)} + q\Pi_{(2,1)} \\ 2\Pi_{(2,0)} = 2q\Pi_{(3,0)} + q\Pi_{(3,1)} \\ \vdots \\ (m+1)\Pi_{(m+1,0)} = (m+1)q\Pi_{(m+2,0)} + q\Pi_{(m+2,1)} \\ (m+2)\Pi_{(m+2,0)} = (m+2)q\Pi_{(m+3,0)} + q\Pi_{(m+3,1)} + p\Pi_{(m+1,0)} \\ \vdots \\ n\Pi_{(n,0)} = nq\Pi_{(n+1,0)} + q\Pi_{(n+1,1)} + (n-m-1)p\Pi_{(n-1,0)} \\ \vdots \end{cases} \tag{4}$$

$$\begin{cases} 2\Pi_{(2,1)} = q\Pi_{(3,1)} + 2q\Pi_{(3,2)} + p\Pi_{(1,0)} + p\Pi_{(1,0)} \\ 3\Pi_{(3,1)} = 2q\Pi_{(4,1)} + 2q\Pi_{(4,2)} + 2p\Pi_{(2,0)} \\ \vdots \\ (m+1)\Pi_{(m+1,1)} = mq\Pi_{(m+2,1)} + 2q\Pi_{(m+2,2)} + mp\Pi_{(m,0)} \\ (m+2)\Pi_{(m+2,1)} = (m+1)q\Pi_{(m+3,1)} + 2q\Pi_{(m+3,2)} + mp\Pi_{(m+1,0)} + p\Pi_{(m+1,1)} \\ \vdots \\ n\tilde{P}_{(n,1)} = (n-1)q\Pi_{(n+1,1)} + 2q\Pi_{(n+1,2)} + mp\Pi_{(n-1,0)} + (n-m-1)p\Pi_{(n-1,1)} \\ \vdots \end{cases} \tag{5}$$

$$\vdots$$

$$\begin{cases} m\Pi_{(m,m-1)} = q\Pi_{(m+1,m-1)} + mq\Pi_{(m+1,m)} + (m-1)p\Pi_{(m-1,m-2)} + p\left[\sum_{i=0}^{m-2}\Pi_{(m-1,i)}\right] \\ (m+1)\Pi_{(m+1,m-1)} = 2q\Pi_{(m+2,m-1)} + mq\Pi_{(m+2,m)} + mp\Pi_{(m,m-2)} \\ (m+2)\Pi_{(m+2,m-1)} = 3q\Pi_{(m+3,m-1)} + mq\Pi_{(m+3,m)} + mp\Pi_{(m+1,m-2)} + (m+2-m-1)p\Pi_{(m+1,m-1)} \\ \vdots \\ n\Pi_{(n,m-1)} = (n-m+1)q\Pi_{(n+1,m-1)} + mq\Pi_{(n+1,m)} + mp\Pi_{(n-1,m-2)} + (n-m-1)p\Pi_{(n-1,m-1)} \\ \vdots \end{cases}$$

$$\begin{cases} (m+1)\Pi_{(m+1,m)} = q\Pi_{(m+2,m)} + (m+1)q\Pi_{(m+2,m+1)} + mp\Pi_{(m,m-1)} + p\left[\sum_{i=0}^{m-1}\Pi_{(m,i)}\right] \\ (m+2)\Pi_{(m+2,m)} = 2q\Pi_{(m+3,m)} + (m+1)q\Pi_{(m+3,m+1)} + mp\Pi_{(m+1,m-1)} + p\Pi_{(m+1,m)} + p\left[\sum_{i=0}^{m}\Pi_{(m+1,i)}\right] \\ \vdots \\ n\Pi_{(n,m)} = (n-m)q\Pi_{(n+1,m)} + (m+1)q\Pi_{(n+1,m+1)} + mp\Pi_{(n-1,m-1)} + (n-m-1)p\Pi_{(n-1,m)} + p\left[\sum_{i=0}^{n-2}\Pi_{(n-1,i)}\right] \\ \vdots \end{cases} \tag{7}$$

$$\begin{cases} (r+1)\Pi_{(r+1,r)} = q\Pi_{(r+2,r)} + (r+1)q\Pi_{(r+2,r+1)} + mp\Pi_{(r,r-1)} \\ (r+2)\Pi_{(r+2,r)} = 2q\Pi_{(r+3,r)} + (r+1)q\Pi_{(r+3,r+1)} + mp\Pi_{(r+1,r-1)} + (r+2-m-1)p\tilde{P}_{(r+1,r)} \\ \vdots \\ n\Pi_{(n,r)} = (n-r)q\Pi_{(n+1,r)} + (r+1)q\Pi_{(n+1,r+1)} + mp\Pi_{(n-1,r-1)} + (n-m-1)p\Pi_{(n-1,r)} \\ \vdots \end{cases} \tag{8}$$

$r \geq m+1$

### 2.3 Steady state degree distribution equations

Let $K$ be the steady state degree distribution [15,23,24], and $\Pi(k)$ be the probability distribution of $K$, that is,

$$\Pi(k)=P\{K=k\}=\lim_{t\to+\infty}\sum_{i\geq k+1}\tilde{P}_{(i,k)}(t)=\sum_{i\geq k+1}\Pi_{(i,k)} \tag{9}$$

Combining Eq.(9) and steady state transformation equations (4)-(8), we can obtain the steady state degree distribution equations as follows:

$$\begin{cases}
(q+mp)\Pi(0)=q\Pi(1)+q\Pi_{(1,0)}+\sum_{i=1}^{m-1}(m-i)\,p\Pi_{(i,0)} \\
(2q+mp)\Pi(1)=2q\Pi(2)+mp\Pi(0)+p\Pi_{(1,0)}-\sum_{i=1}^{m-1}(m-i)\,p\Pi_{(i,0)}+\sum_{i=2}^{m-1}(m-i)\,p\Pi_{(i,1)} \\
\vdots \\
(mq+mp)\Pi(m-1)=mq\Pi(m)+mp\Pi(m-2)+p\sum_{i=0}^{m-3}\Pi_{(m-1,i)} \\
\left[(m+1)q+mp\right]\Pi(m)=(m+1)q\Pi(m+1)+mp\Pi(m-1)+p-p\sum_{i=1}^{m-1}\Pi_N(i) \\
\left[(m+2)q+mp\right]\Pi(m+1)=(m+2)q\Pi(m+2)+mp\Pi(m) \\
\vdots \\
\left[(r+1)q+mp\right]\Pi(r)=(r+1)q\Pi(r+1)+mp\Pi(r-1) \\
\vdots
\end{cases} \tag{10}$$

## 3. Degree Distribution of RBDN with Network Size Decline

Let $N(t)$ be the number of nodes in RBDN at time $t$ and $N(0)=m+1$. Let $N$ be the steady state network size, and $\Pi_N(n)$ be the probability distribution of $N$, that is,

$$\Pi_N(n)=P\{N=n\}=\lim_{t\to+\infty}P\{N(t)=n\}=\sum_{k=0}^{n-1}\Pi_{(n,k)} \quad n\geq 1 \tag{11}$$

Considering $\{N(t),t\geq 0\}$ is an one dimensional random walk with a left bound 1, we have

$$\Pi_N(n)=\frac{q-p}{q}\left(\frac{p}{q}\right)^{n-1} \quad n\geq 1 \tag{12}$$

As shown in Eq. (10)，in the case of *m*=1, 2, we only need to calculate $\Pi_{(1,0)}$ before the probability generation function method is employed for the exact solution of the degree distribution [22], in which $\Pi_{(1,0)}$ can be obtained directly as follows:

$$\Pi_{(1,0)} = \Pi_N(1) = \frac{q-p}{q} \tag{13}$$

However, in the case of $m \geqslant 3$, $\Pi_{(i,k)}$ is required for the calculation, rather than $\Pi_{(1,0)}$ which is a special cases. Obviously, $\Pi_{(i,k)}$ is much more difficult to obtain. Thus in the following section, we focus on the calculation of $\Pi_{(i,k)}$.

## 3.1 Calculation of $\Pi_{(i,k)}$

To obtain $\Pi_{(i,k)}$, $\tilde{e}_n$ is introduced，i.e.

$$\tilde{e}_n = \sum_{k=0}^{n-1} k \cdot \Pi_{(n,k)} \tag{14}$$

Since

$$\begin{aligned} E[K] &= \sum_k k\Pi(k) = \sum_k \sum_{i \geq k+1} k\Pi_{(i,k)} \\ &= \sum_{i=1}^{+\infty} \sum_{k=0}^{i-1} k\Pi_{(i,k)} = \sum_{i=1}^{+\infty} \tilde{e}_i \end{aligned} \tag{15}$$

$\tilde{e}_n$ denotes the contribution of the network with *n* nodes to the average degree $E[K]$. From Eq.(14) and the steady state transformation equations (4-8), we can obtain the equations about $\tilde{e}_i \ (i = 1, 2, \cdots)$, as follows:

$$\begin{cases} n\tilde{e}_n = (n-1)p\tilde{e}_{n-1} + (n-1)q\tilde{e}_{n+1} + 2(n-1)p\Pi_N(n-1) & 2 \leq n \leq m \\ n\tilde{e}_n = (n-1)p\tilde{e}_{n-1} + (n-1)q\tilde{e}_{n+1} + 2mp\Pi_N(n-1) & n \geq m+1 \end{cases} \tag{16}$$

where

$$\tilde{e}_1 = 0 \tag{17}$$

To sum up the first $n\ (n > m)$ items in Eq.(16), we can get

$$2q\sum_{i=2}^{n+1} \tilde{e}_i = -(n+1)p\tilde{e}_{n+1} + nq\tilde{e}_{n+2} + 2p\sum_{i=1}^{m-1} i\Pi_N(i) + 2mp\sum_{i=m}^{n} \Pi_N(i) \tag{18}$$

Note

$$\tilde{e}_i \geq 0 \quad and \quad \sum_{i=2}^{+\infty} \tilde{e}_i < 2m \tag{19}$$

then we have

$$\lim_{n\to+\infty} n\tilde{e}_n = 0 \tag{20}$$

From Eq.(18), we have

$$\sum_{i=2}^{+\infty} \tilde{e}_i = \frac{p}{q}\left[\sum_{i=1}^{m-1} i\Pi_N(i) + m\sum_{i=m}^{+\infty} \Pi_N(i)\right] = \frac{p}{q}\left[m - \frac{q-p}{q}\sum_{i=1}^{m-1}(m-i)\left(\frac{p}{q}\right)^{i-1}\right] \tag{21}$$

Then generation function for Eq. (16) can be rewritten as

$$T(x) = \sum_{i=0}^{+\infty} \tilde{e}_{2+i} x^i \tag{22}$$

satisfying

$$\begin{cases} T'(x) = \dfrac{2(1-px)}{(1-x)(q-px)} T(x) - \dfrac{2p(q-p)}{q^2} \dfrac{\left(q^2 + pqx + p^2x^2\right)}{(1-x)(q-px)^2} \\ T(1) = \sum_{i=2}^{+\infty} \tilde{e}_i = \dfrac{p}{q}\left[m - \dfrac{q-p}{q}\sum_{i=1}^{m-1}(m-i)\left(\dfrac{p}{q}\right)^{i-1}\right] \end{cases} \tag{23}$$

Solving the differential equation (23), $\tilde{e}_i \ (i \geq 2)$ can be obtained. Combining Eqs. (14) and (4)-(8)，we have

$$\begin{cases} \Pi_{(2,1)} = \tilde{e}_2 \\ \Pi_{(3,1)} + 2\Pi_{(3,2)} = \tilde{e}_3 \\ \quad \vdots \\ \Pi_{(2,0)} + \Pi_{(2,1)} = \Pi_N(2) \\ \Pi_{(3,0)} + \Pi_{(3,1)} + \Pi_{(3,2)} = \Pi_N(3) \\ \quad \vdots \\ 2q\Pi_{(3,0)} + q\Pi_{(3,1)} = 2\Pi_{(2,0)} \\ q\Pi_{(3,1)} + 2q\Pi_{(3,2)} = 2\Pi_{(2,1)} - p\Pi_{(1,0)} - p\Pi_{(1,0)} \\ \quad \vdots \end{cases} \tag{24}$$

Then $\Pi_{(i,k)}$ can be solved by Eq. (24).

## 3.2 Exact solutions of the degree distributions for *m*=3, 0<*p*<1/2,

Once $\Pi_{(i,k)}$ is obtained, probability generating function approach can be employed for Eq.(10) to obtain the steady state degree distribution $\Pi(k)$. In this section, taking $m$=3 as an example, we explain how this approach is used. From Eq. (10), we can obtain the degree distribution equations of RBDN with 0<p<1/2, *m*=3 as follows:

$$\begin{cases} (q+3p)\Pi(0) = q\Pi(1) + (2p+q)\Pi_{(1,0)} + p\Pi_{(2,0)} \\ (2q+3p)\Pi(1) = 2q\Pi(2) + 3p\Pi(0) - p\Pi_{(1,0)} - p\Pi_{(2,0)} + p\Pi_{(2,1)} \\ (3q+3p)\Pi(2) = 3q\Pi(3) + 3p\Pi(1) + p\Pi_{(2,0)} \\ (4q+3p)\Pi(3) = 4q\Pi(4) + 3p\Pi(2) + p - p\Pi_N(1) - p\Pi_N(2) \\ (5q+3p)\Pi(4) = 5q\Pi(5) + 3p\Pi(3) \\ \quad\vdots \\ \left[(r+1)q+3p\right]\Pi(r) = (r+1)q\Pi(r+1) + 3p\Pi(r-1) \\ \quad\vdots \end{cases} \tag{25}$$

From Eq. (25) we may find that $\Pi_{(2,0)}$ and $\Pi_{(2,1)}$ are needed before we solve the degree distribution.

In the case of *m* =3, Eq. (16) can be rewritten as

$$\begin{cases} n\tilde{e}_n = (n-1)p\tilde{e}_{n-1} + (n-1)q\tilde{e}_{n+1} + 2(n-1)p\Pi_N(n-1) & 2 \le n \le 3 \\ n\tilde{e}_n = (n-1)p\tilde{e}_{n-1} + (n-1)q\tilde{e}_{n+1} + 6p\Pi_N(n-1) & n \ge 4 \end{cases} \tag{26}$$

Here we introduce the following generating function

$$T(x) = \sum_{i=0}^{+\infty} \tilde{e}_{2+i} x^i \tag{27}$$

Combining Eqs.(26), we have:

$$\begin{cases} T'(x) = \dfrac{2(1-px)}{(1-x)(q-px)} T(x) - \dfrac{2p(q-p)}{q^2} \dfrac{(q^2+pqx+p^2x^2)}{(1-x)(q-px)^2} \\ T(1) = \dfrac{p(1-pq)}{q^3} \end{cases} \tag{28}$$

Solving the differential equation (28), we have

$$T(x) = \frac{2p(q-p)}{q^2} \frac{(q-px)^{\frac{2p}{q-p}}}{(1-x)^{\frac{2q}{q-p}}} \int_x^1 \frac{(q^2+pqt+p^2t^2)(1-t)^{\frac{1}{q-p}}}{(q-pt)^{\frac{2q}{q-p}}} dt \tag{29}$$

Thus

$$\tilde{e}_2 = T(0) = \frac{2p(q-p)}{q^2} q^{\frac{2p}{q-p}} \int_0^1 \frac{(q^2+pqt+p^2t^2)(1-t)^{\frac{1}{q-p}}}{(q-pt)^{\frac{2q}{q-p}}} dt \tag{30}$$

Let

$$\frac{1-t}{q-pt} = y \tag{31}$$

then

$$t=\frac{1-qy}{1-py}\,,\quad dt=-\frac{q-p}{\left(1-py\right)^2}dy \tag{32}$$

So, we have

$$\begin{aligned}\tilde{e}_2&=\frac{2p\left(q-p\right)^2}{q^2}q^{\frac{2p}{q-p}}\left[p^2\left(q-p\right)\int_0^{\frac{1}{q}}\frac{y^{\frac{2q}{q-p}+1}}{\left(1-py\right)^3}dy-p\left(1+p\right)\int_0^{\frac{1}{q}}\frac{y^{\frac{2q}{q-p}}}{\left(1-py\right)^2}dy+\frac{1-pq}{q-p}\int_0^{\frac{1}{q}}\frac{y^{\frac{1}{q-p}}}{1-py}dy\right]\\&=\frac{2p\left(q-p\right)}{q^2}q^{\frac{2p}{q-p}}\left[\left(1+2q+3q^2\right)\int_0^{\frac{1}{q}}\frac{y^{\frac{1}{q-p}}}{1-py}dy-\frac{1}{2}\left(2+3q\right)q^{-\frac{1}{q-p}}\right]\end{aligned} \tag{33}$$

$$=\frac{p\left(q-p\right)}{q^2}q^{\frac{2p}{q-p}}\left[2\left(1+2q+3q^2\right)p^{-\frac{2q}{q-p}}\int_0^{\frac{p}{q}}\frac{x^{\frac{1}{q-p}}}{1-x}dx-\left(2+3q\right)q^{-\frac{1}{q-p}}\right]$$

For

$$\begin{cases}\Pi_{(2,1)}=\tilde{e}_2\\ \Pi_{(2,0)}+\Pi_{(2,1)}=\Pi_N\left(2\right)\end{cases} \tag{34}$$

Rewriting (34), we have

$$\begin{cases}\Pi_{(2,1)}=\tilde{e}_2\\ \Pi_{(2,0)}=\Pi_N\left(2\right)-\Pi_{(2,1)}\end{cases} \tag{35}$$

To solve $\Pi\left(k\right)$, let the probability generating function be

$$G\left(x\right)=\sum_{i=0}^{+\infty}\Pi\left(i\right)\cdot x^i\,,\quad G\left(1\right)=\sum_{i=0}^{+\infty}\Pi\left(i\right)=1 \tag{36}$$

According to Eq.(25) and Eq.(35), we can get

$$\begin{aligned}G'\left(x\right)=&\frac{3px-q-3p}{q\left(x-1\right)}G\left(x\right)+\frac{p\left(1-\Pi_N\left(1\right)-\Pi_N\left(2\right)\right)}{q\left(x-1\right)}x^3+\frac{p\Pi_{(2\ ,)0}}{q\left(x-1\right)}x^2\\&+\frac{p\left(\Pi_{(2,1)}-\Pi_{(2,0)}-\Pi_N\left(1\right)\right)}{q\left(x-1\right)}x+\frac{\Pi_N\left(1\right)+p\Pi_N\left(1\right)+p\Pi_{(2,0)}}{q\left(x-1\right)}\end{aligned} \tag{37}$$

Solving the Eq.(37), we can obtain

$$G(x)=\frac{e^{3px/q}}{1-x}\int_x^1\left[\begin{array}{l}\frac{p\left(1-\Pi_N(1)-\Pi_N(2)\right)}{q}t^3+\frac{p\Pi_{(2,0)}}{q}t^2\\+\frac{p\left(\Pi_{(2,1)}-\Pi_{(2,0)}-\Pi_N(1)\right)}{q}t+\frac{(1+p)\Pi_N(1)+p\Pi_{(2,0)}}{q}\end{array}\right]e^{-3pt/q}dt$$

$$=\left(\frac{6q^2-2q+6}{27pq}-\frac{2\tilde{e}_2q^2}{27p^2}\right)e^{-3p/q}\left[\sum_{k=0}^{+\infty}x^k\sum_{i=k+1}^{+\infty}\frac{1}{i!}\left(\frac{3p}{q}\right)^i\right]$$
$$-\left(\frac{1+2p}{9q^2}+\frac{3p-2q}{9p}\tilde{e}_2\right)-\left(\frac{2p}{3q}-\frac{1}{3}\tilde{e}_2\right)x-\frac{p^2}{3q^2}x^2 \tag{38}$$

So， the degree distributions of RBDN for 0<*p*<1/2, *m*=3 are as follows

$$\Pi(k)=\begin{cases}c\cdot e^{-3p/q}\sum_{i=k+1}^{+\infty}\frac{1}{i!}\left(\frac{3p}{q}\right)^i-a_0, & k=0\\ c\cdot e^{-3p/q}\sum_{i=k+1}^{+\infty}\frac{1}{i!}\left(\frac{3p}{q}\right)^i-a_1, & k=1\\ c\cdot e^{-3p/q}\sum_{i=k+1}^{+\infty}\frac{1}{i!}\left(\frac{3p}{q}\right)^i-a_2, & k=2\\ c\cdot e^{-3p/q}\sum_{i=k+1}^{+\infty}\frac{1}{i!}\left(\frac{3p}{q}\right)^i, & k\geq 3\end{cases} \tag{39}$$

where

$$\begin{cases}a_0=\frac{1+2p}{9q^2}+\frac{3p-2q}{9p}\tilde{e}_2\\ a_1=\frac{2p}{3q}-\frac{1}{3}\tilde{e}_2\\ a_2=\frac{p^2}{3q^2}\\ c=\frac{6q^2-2q+6}{27pq}-\frac{2\tilde{e}_2q^2}{27p^2}\\ \tilde{e}_2=\frac{p(q-p)}{q^2}q^{\frac{2p}{q-p}}\left[2\left(1+2q+3q^2\right)p^{-\frac{2q}{q-p}}\int_0^{\frac{p}{q}}\frac{x^{\frac{1}{q-p}}}{1-x}dx-(2+3q)q^{-\frac{1}{q-p}}\right]\end{cases} \tag{40}$$

Figures 1 illustrates the exact solutions and simulation results of the degree distributions for 0<*p*<1/2, $m=3$, where the horizontal and vertical ordinates denote the degree of nodes and the probability respectively. Each simulation number is the average value of 1000 simulation results for $t=10000$. As shown in Fig. 1, the exact solutions match perfectly with the numerical solutions, verifying the correctness of our exact solutions.

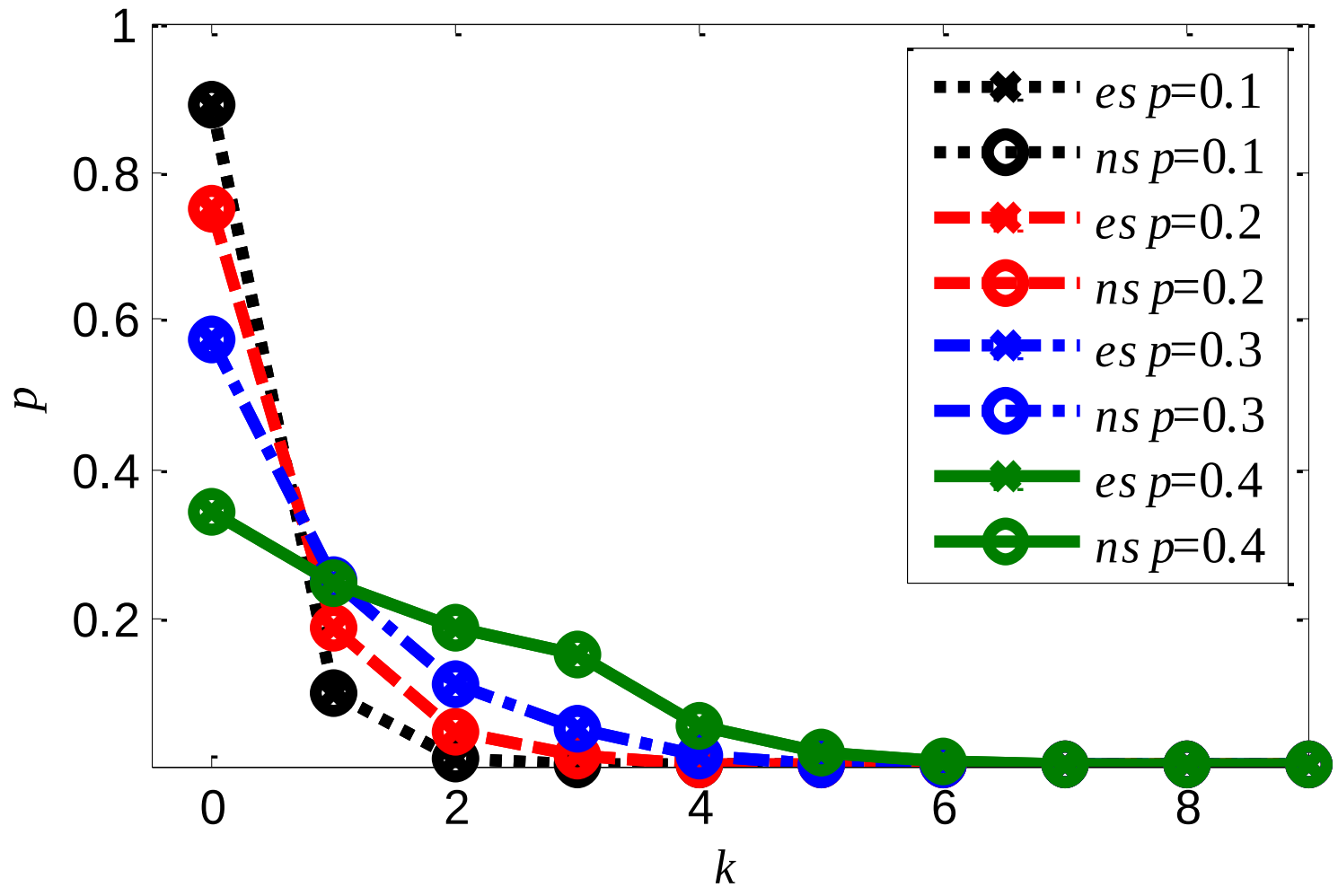

Fig. 1 Exact solutions vs. numerical solutions ($t$=10000): degree distributions of RBDN for $p$<1/2, $m$=3

## 4. Poisson Tail

By Eq. (39), we can conclude that RBDN with 0<$p$<1/2, for the large $k$ $\left(k \geq m+1\right)$, the degree distribution of RBDN exhibits a backward accumulation form of Poisson distribution, that is,

$$\Pi\left(k\right)=\alpha\cdot\sum_{r=k+1}^{+\infty}\frac{\lambda^{r}}{r!} \quad k\geq m+1 \tag{41}$$

where $\alpha$ is a positive constant and $\lambda=\frac{mp}{q}$. For sufficient large $k$, we have

$$\Pi\left(k\right)=\alpha\cdot\sum_{r=k+1}^{+\infty}\frac{\lambda^{r}}{r!}\sim\alpha\frac{\lambda^{k+1}}{\left(k+1\right)!} \tag{42}$$

Thus in the case of 0<$p$<1/2, the degree distribution of RBDN exhibits a Poisson tail. Here we employ the same approach as in [22] to verify this Poisson tail for the degree distribution.

Let

$$r\left(k\right)=\frac{\Pi\left(k\right)}{\Pi\left(k\text{-}1\right)} \tag{43}$$

Then using Eq.(42), we have

$$r\left(k\right)\sim\frac{\lambda}{k}; \quad \ln r\left(k\right)\sim\ln\lambda-\ln k \tag{44}$$

that is, $r\left(k\right)$ is a line with slope -1 for large $k$ in the two-logarithm axis diagram.

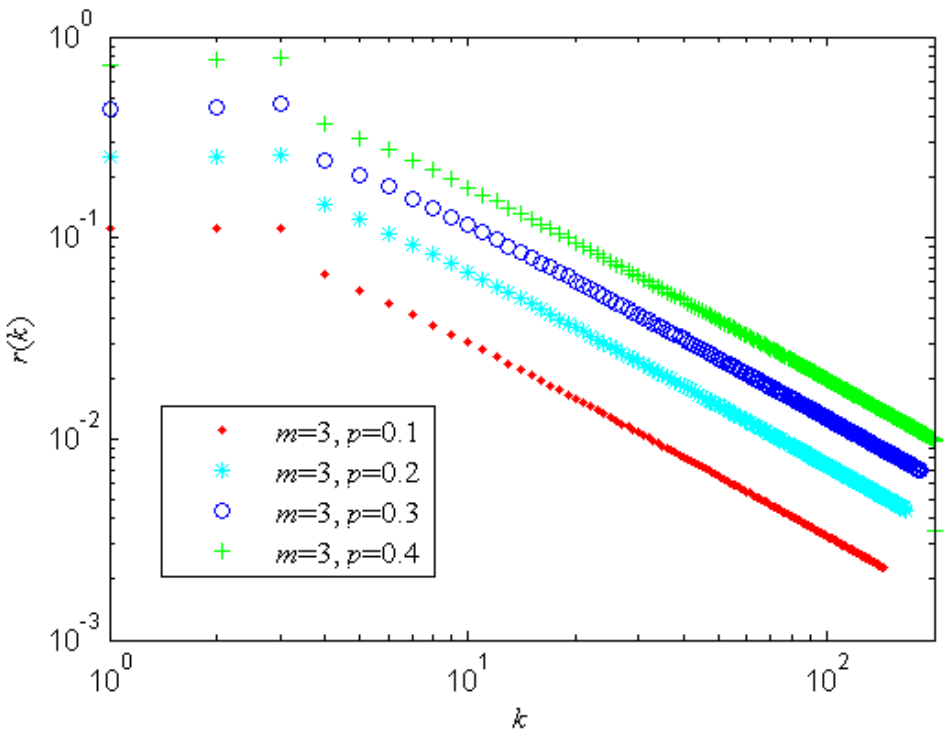

Fig.2-a

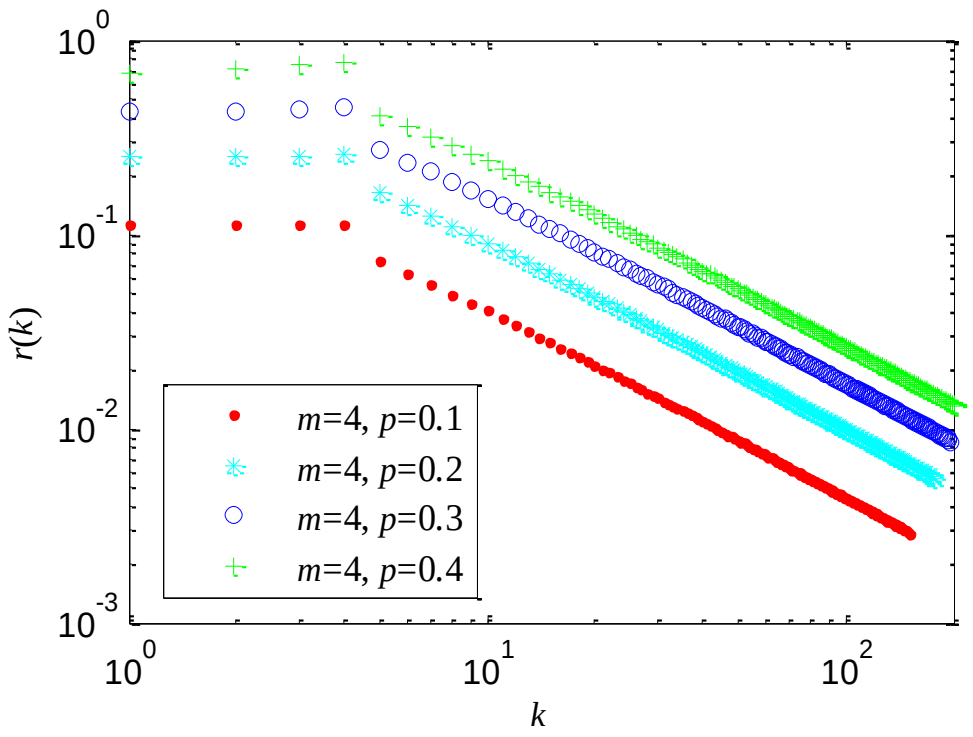

Fig.2-b

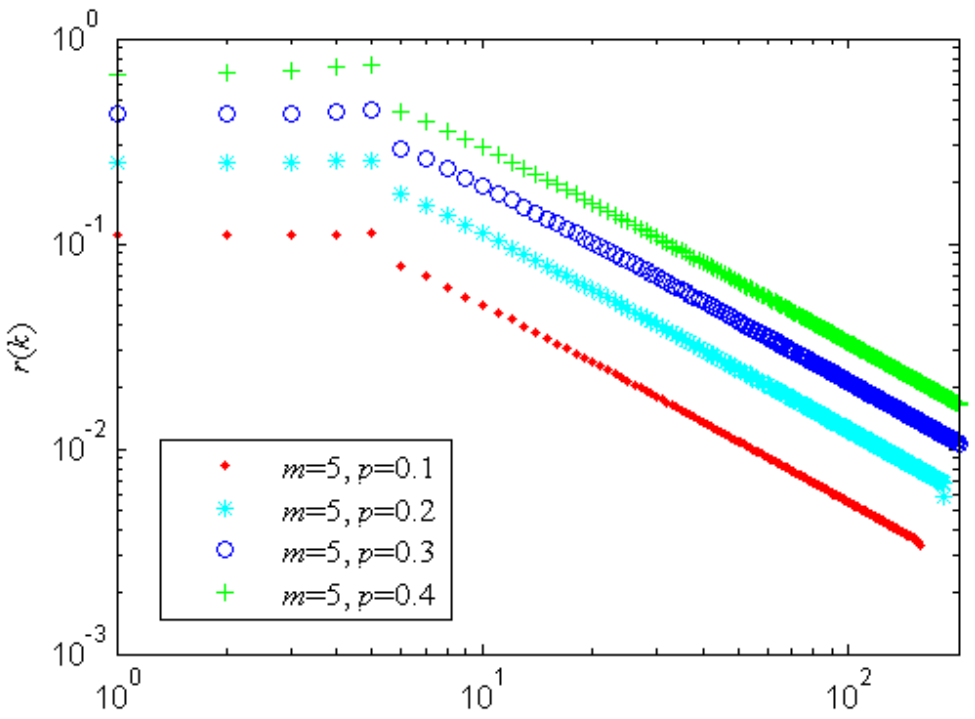

Fig.2-c

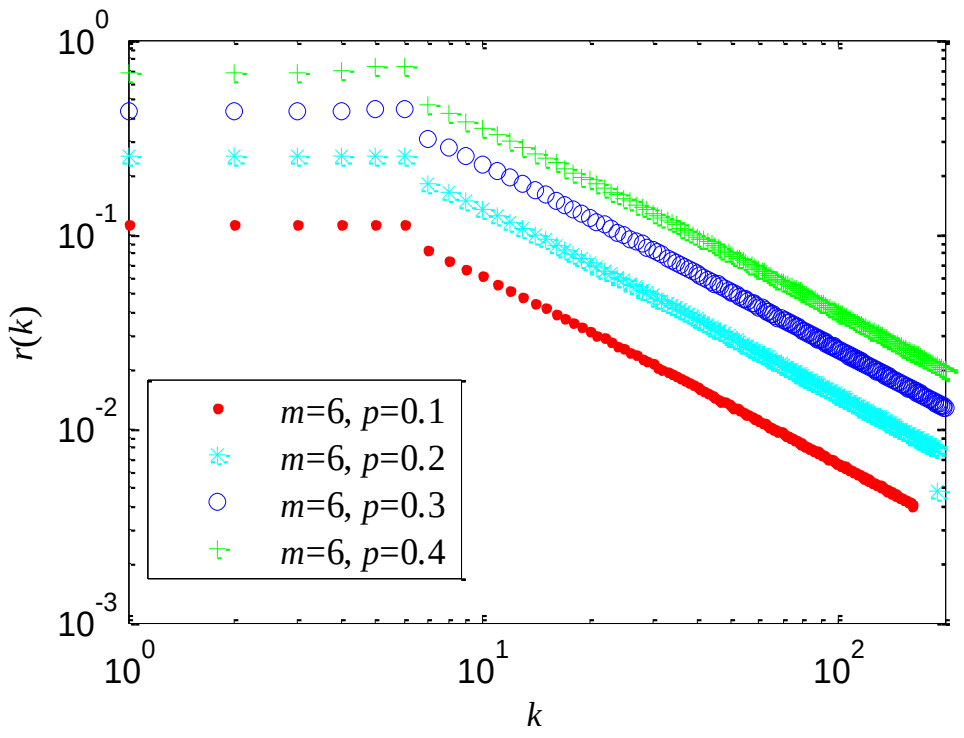

Fig.2-d

Fig.2 The Poisson tail of RBDN with 0<p<1/2

Fig.2 illustrates the Poisson tails of the degree distributions for various $p$ ($0<p<1/2$). As we can see, in the case $k \geq m+1$, the slopes of lines tend to be -1, showing the Poisson tails for the degree distributions of RBDN.

## 5. Conclusion

In this paper, we provide a general approach to obtain the exact solutions of the degree distributions in the case of $m \geqslant 3$, $0<p<1/2$. Especially, taking $m=3$ as an example, we explain the detailed solving process, in which computer simulation is used to verify our degree distribution solutions. In addition, the characteristics of the degree distribution are discussed. Our findings suggest that the degree distributions will exhibit Poisson tail property for the declining RBDN.

### Acknowledgments

This research was financially supported by the National Natural Science Foundation of China (No. 61273015) and the China Scholarship Council.

## Appendix 1: State transformation equations

The state transformation equations of $NK(t)$ are as follows:

$$\begin{cases} \tilde{P}_{(1,0)}(t+1) = q\tilde{P}_{(1,0)}(t) + q\tilde{P}_{(2,0)}(t) + q\tilde{P}_{(2,1)}(t) \\ 2\tilde{P}_{(2,0)}(t+1) = 2q\tilde{P}_{(3,0)}(t) + q\tilde{P}_{(3,1)}(t) \\ \qquad \vdots \\ (m+1)\tilde{P}_{(m+1,0)}(t+1) = (m+1)q\tilde{P}_{(m+2,0)}(t) + q\tilde{P}_{(m+2,1)}(t) \\ (m+2)\tilde{P}_{(m+2,0)}(t+1) = (m+2)q\tilde{P}_{(m+3,0)}(t) + q\tilde{P}_{(m+3,1)}(t) + p\tilde{P}_{(m+1,0)}(t) \\ \qquad \vdots \\ n\tilde{P}_{(n,0)}(t+1) = nq\tilde{P}_{(n+1,0)}(t) + q\tilde{P}_{(n+1,1)}(t) + (n-m-1)p\tilde{P}_{(n-1,0)}(t) \\ \qquad \vdots \end{cases} \tag{45}$$

$$\begin{cases} 2\tilde{P}_{(2,1)}(t+1) = q\tilde{P}_{(3,1)}(t) + 2q\tilde{P}_{(3,2)}(t) + p\tilde{P}_{(1,0)}(t) + p\tilde{P}_{(1,0)}(t) \\ 3\tilde{P}_{(3,1)}(t+1) = 2q\tilde{P}_{(4,1)}(t) + 2q\tilde{P}_{(4,2)}(t) + 2p\tilde{P}_{(2,0)}(t) \\ \qquad \vdots \\ (m+1)\tilde{P}_{(m+1,1)}(t+1) = mq\tilde{P}_{(m+2,1)}(t) + 2q\tilde{P}_{(m+2,2)}(t) + mp\tilde{P}_{(m,0)}(t) \\ (m+2)\tilde{P}_{(m+2,1)}(t+1) = (m+1)q\tilde{P}_{(m+3,1)}(t) + 2q\tilde{P}_{(m+3,2)}(t) + mp\tilde{P}_{(m+1,0)}(t) + p\tilde{P}_{(m+1,1)}(t) \\ \qquad \vdots \\ n\tilde{P}_{(n,1)}(t+1) = (n-1)q\tilde{P}_{(n+1,1)}(t) + 2q\tilde{P}_{(n+1,2)}(t) + mp\tilde{P}_{(n-1,0)}(t) + (n-m-1)p\tilde{P}_{(n-1,1)}(t) \\ \qquad \vdots \end{cases} \tag{46}$$

$$\vdots$$

$$\begin{cases} m\tilde{P}_{(m,m-1)}(t+1) = q\tilde{P}_{(m+1,m-1)}(t) + mq\tilde{P}_{(m+1,m)}(t) + (m-1)p\tilde{P}_{(m-1,m-2)}(t) + p\left[\sum_{i=0}^{m-2}\tilde{P}_{(m-1,i)}(t)\right] \\ (m+1)\tilde{P}_{(m+1,m-1)}(t+1) = 2q\tilde{P}_{(m+2,m-1)}(t) + mq\tilde{P}_{(m+2,m)}(t) + mp\tilde{P}_{(m,m-2)}(t) \\ (m+2)\tilde{P}_{(m+2,m-1)}(t+1) = 3q\tilde{P}_{(m+3,m-1)}(t) + mq\tilde{P}_{(m+3,m)}(t) + mp\tilde{P}_{(m+1,m-2)}(t) \\ \qquad + (m+2-m-1)p\tilde{P}_{(m+1,m-1)}(t) \\ \qquad \vdots \\ n\tilde{P}_{(n,m-1)}(t+1) = (n-m+1)q\tilde{P}_{(n+1,m-1)}(t) + mq\tilde{P}_{(n+1,m)}(t) + mp\tilde{P}_{(n-1,m-2)}(t) \\ \qquad + (n-m-1)p\tilde{P}_{(n-1,m-1)}(t) \\ \qquad \vdots \end{cases} \tag{47}$$

$$\begin{cases} (m+1)\tilde{P}_{(m+1,m)}(t+1) = q\tilde{P}_{(m+2,m)}(t) + (m+1)q\tilde{P}_{(m+2,m+1)}(t) + mp\tilde{P}_{(m,m-1)}(t) \\ \qquad + p\left[\sum_{i=0}^{m-1}\tilde{P}_{(m,i)}(t)\right] \\ (m+2)\tilde{P}_{(m+2,m)}(t+1) = 2q\tilde{P}_{(m+3,m)}(t) + (m+1)q\tilde{P}_{(m+3,m+1)}(t) + mp\tilde{P}_{(m+1,m-1)}(t) \\ \qquad + p\tilde{P}_{(m+1,m)}(t) + p\left[\sum_{i=0}^{m}\tilde{P}_{(m+1,i)}(t)\right] \\ \qquad \vdots \\ n\tilde{P}_{(n,m)}(t+1) = (n-m)q\tilde{P}_{(n+1,m)}(t) + (m+1)q\tilde{P}_{(n+1,m+1)}(t) + mp\tilde{P}_{(n-1,m-1)}(t) \\ \qquad + (n-m-1)p\tilde{P}_{(n-1,m)}(t) + p\left[\sum_{i=0}^{n-2}\tilde{P}_{(n-1,i)}(t)\right] \\ \qquad \vdots \end{cases} \tag{48}$$

$$\begin{cases} (r+1)\tilde{P}_{(r+1,r)}(t+1) = q\tilde{P}_{(r+2,r)}(t) + (r+1)q\tilde{P}_{(r+2,r+1)}(t) + mp\tilde{P}_{(r,r-1)}(t) \\ (r+2)\tilde{P}_{(r+2,r)}(t+1) = 2q\tilde{P}_{(r+3,r)}(t) + (r+1)q\tilde{P}_{(r+3,r+1)}(t) + mp\tilde{P}_{(r+1,r-1)}(t) \\ \qquad + (r+2-m-1)p\tilde{P}_{(r+1,r)}(t) \\ \qquad \vdots \qquad\qquad r \ge m+1 \\ n\tilde{P}_{(n,r)}(t+1) = (n-r)q\tilde{P}_{(n+1,r)}(t) + (r+1)q\tilde{P}_{(n+1,r+1)}(t) + mp\tilde{P}_{(n-1,r-1)}(t) \\ \qquad + (n-m-1)p\tilde{P}_{(n-1,r)}(t) \\ \qquad \vdots \end{cases} \tag{49}$$